\documentclass[10pt,a4paper]{article}
\newcommand\abs[1]{\left|#1\right|}

\usepackage{multicol}

\usepackage{titlesec}

\titleformat{\chapter}{\normalfont\huge}{\thechapter.}{20pt}{\huge\textbf}

\usepackage[english]{babel} 
\usepackage[T1]{fontenc}
\usepackage{makeidx}
\usepackage{appendix}
\usepackage{amsfonts}
\usepackage{amsmath, relsize}
\usepackage{amsmath}
\usepackage{slashed}
\usepackage{amssymb}
\usepackage{graphicx}
\usepackage{mathrsfs}
\usepackage{amsthm}
\usepackage{amssymb}
\usepackage{verbatim}
\usepackage{booktabs}
\usepackage{pxfonts}
\usepackage{hyperref}
\hypersetup{
     colorlinks   = true,
     citecolor    = blue,
     linkcolor   = blue
}

\usepackage[top=2.5cm, bottom=2.5cm, left=2.0cm, right=1.8cm]{geometry}

\usepackage{caption}

\usepackage{float} 

\author{Chiara Gastaldi}
\date{\today}

\begin{document}


\begin{center}
\begin{large}
\textbf{The contribution of NLO and LPM corrections to thermal dilepton emission\\ in heavy ion collisions}\\
\end{large}
\vspace{0.4cm}
Yannis Burnier, Chiara Gastaldi\\
\vspace{0.4cm}
\small{\em
  Institut de Th\'eorie des Ph\'enom\`enes Physiques,\\
   \'Ecole Polytechnique
F\'ed\'erale de Lausanne, CH-1015 Lausanne, Switzerland
}\\
\vspace{0.1cm}
(\today)
\end{center}

\vskip 0.2cm


\begin{abstract}

Recently lots of efforts have been made to obtain the next to leading order and Landau-Pomeranchuk-Migdal corrections to the thermal dilepton emission rate in perturbative QCD. Here we apply these results to the plasma created in heavy ion collisions and see wether these corrections improve the comparison between theoretical calculations and experimental results for the invariant mass dependence of the dilepton emission rate. In particular, we simulate the quark-gluon plasma produced at RHIC and LHC using a 2+1-dimensional viscous hydro model. We compare our results to STAR experiment and comment on the need for a non-perturbative determination of the dilepton rate at low invariant mass.
\end{abstract}

\begin{multicols}{2}
\section{Introduction}
The theoretical study of the quark-gluon plasma (QGP)  \cite{QGP1,QGP} is supported by experiments based on ultra-relativistic heavy ion collisions (HIC): gold ions are used at
 RHIC (Relativistic Heavy Ion Collider) at the Brookhaven National Laboratory and lead ions are used at LHC (Large Hadron Collider) at CERN.
 
Excellent probes to study the QGP are  photons and dileptons pairs \cite{W}, in fact, as they interact electromagnetically, they have a small cross section with the strongly interacting matter inside the QGP. Thus  they leave the QGP and reach the detector, carrying information from deep into the plasma  \cite{phenix,star}.
Moreover we prefer dilepton to photons for two reasons: photons are produced from a bigger background of decays \cite{ghiglieri} while, on the other hand, leptons have a non-null invariant mass M, which helps in disentangling various sources \cite{sources}. 

In fact, the dilepton background is also not small: dileptons are produced in every phase of the HIC and  in several types of processes \cite{sources}.
Here we are interested in thermal dileptons \cite{sources} produced by the partonic interactions during the hydrodynamical expansion; these dileptons can tell us about the QGP properties . Thermal dilepton are produced mainly in quark-antiquark annihilation and Compton scattering and their contribution to the spectrum is important in the intermediate invariant mass range, $M\in [0.2,2.5]$ GeV. 

The first type of background we encounter consists in hadronic reactions at early times. They consist in  jet-dilepton conversion from the initial hadronic scattering processes and from the photoproduction processes. These are hard processes that contribute to the spectrum in the high mass range (M>3 GeV).

Secondly particle decays imprint broad peaks in the spectrum, for instance, the contribution from the  decay of open charm $c\bar{c}\rightarrow e^+ e^- X$ is also very important in the intermediate mass range \cite{Vujanovic:2012nq,Vujanovic:2013jpa}.
In the low mass range $0.6<M<1.1$ GeV the decays of vector mesons, i.e. $\rho$ , $\omega$, $\phi$, give a sizeable contribution to the invariant mass spectrum \cite{star,Vujanovic:2012nq,Vujanovic:2013jpa}. Finally below $M<0.2$ pions decays from the hadronic phase dominate.

A big effort has been done to study the thermal dilepton production from the QGP in
 perturbative  QCD: in particular   references \cite{W,strick} provide a complete discussion of the leading order (LO) with inclusion of anisotropic corrections and  \cite{ghiglieri,laine} supply the passage from the LO to the NLO and Landau-Pomeranchuk-Migdal corrections.

In this work we investigate whether higher order corrections in  perturbation theory can improve the comparison between theoretical predictions for the thermal dilepton emission and experimental results. 
 
In section \ref{theo}, we introduce the theoretical background, in particular how to compute the LO dilepton emission rate per unit of 4-volume and per unit of 4-momentum, what is the effect of NLO and LPM corrections to it and how we compute the invariant mass spectrum. 
In section \ref{num}, we explain how we describe the hydrodynamical plasma evolution using SONIC and the details of the numerical computation for the invariant mass spectrum.
In section \ref{res}, we show our results and compare them to experimental result from the STAR experiment at RHIC \cite{star}. We then make analogous computations for LHC and conclude in section \ref{concl}.

\section{Dileptons in heavy ion collisions}
\label{theo}

We recall here the perturbative QCD calculations of the dilepton rate and explain how to use them in the geometry of heavy ion collisions. We use natural units when not stated otherwise and metric signature is (+,-,-,-). In perturbation theory, two distinct expansions are made, one in the electromagnetic coupling, where the LO is sufficient and a second one in the strong coupling. In this work we discuss the validity of this second expansion in the case of the plasma created in heavy ion collisions.

\subsection{Leading order dilepton production rate}
\label{prod_rate}
\begin{figure}[H]
\begin{center}
\includegraphics[scale=0.1]{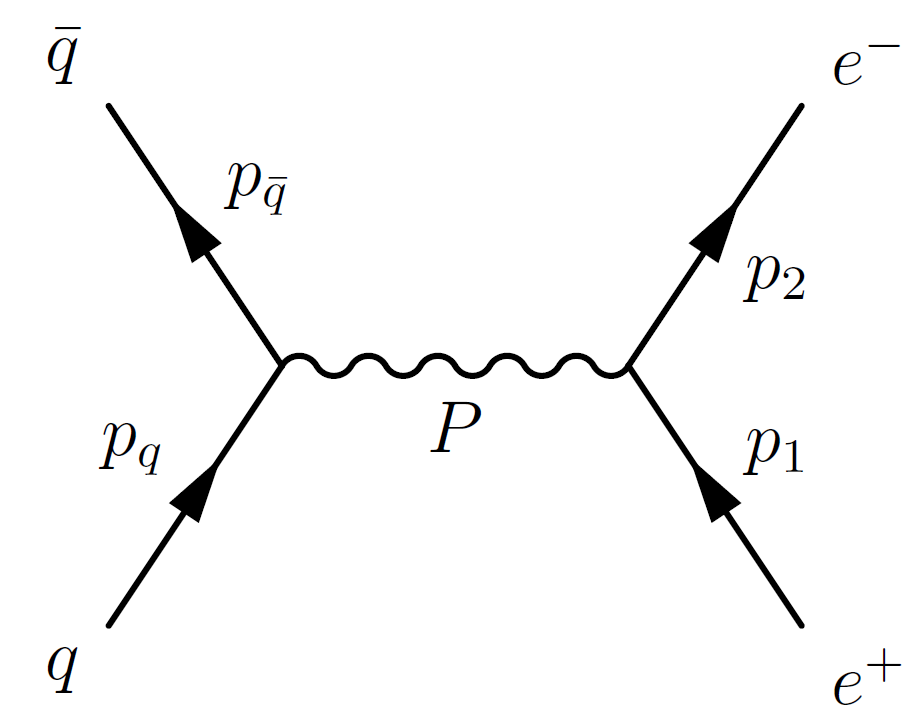}
\vspace{-3mm}
\end{center}
\caption{LO Feynmann diagram for the thermal dilepton production from the QGP.}
\label{feyLO}
\end{figure}

The relation between the dilepton emission rate and thermal expectation values of electromagnetic current correlation function 
\begin{equation}
\label{current}
W^{\mu\nu}=\int d^4x \, e^{-iq\cdot x}\langle J^{\mu}(x)J^{\nu}(0)\rangle,
\end{equation}
is well described  in refs. \cite{W,Weldon:1990iw,book}. Here we briefly summarize the main results for the specific case of a $q \bar{q} \rightarrow e^+e^- $ process, shown in figure \ref{feyLO}.

The number of dileptons produced per unit volume and emitted at a given total momentum $P=(p^0,p^i)$ 
can be expressed trough the dilepton rate $R$:
\begin{equation}
\frac{d N^{\ell \bar \ell}(x,P)}{d^4x d^4P}=\frac{dR(x,P)}{d^4 P},
\end{equation}
which in turn can be calculated form the quark current correlator $W^{\mu\nu}(P)$ as:
\begin{eqnarray}
\frac{dR^{\ell \bar \ell}}{d^{4}P}&=&\sum_{i=1}^{n_f} Q_i^2\frac{\alpha_e^2}{24\pi^3P^2}\left(1+\frac{2m^2}{P^2}\right)\left(1-\frac{4m^2}{P^2}\right)^{\frac{1}{2}}\\&&\quad\quad\times\, \theta \left(P^2-4m^2 \right)W_{\mu}^{\mu}(P).\notag
\end{eqnarray}
where $m$ is the mass of the emitted leptons and $Q_i,\, i=1,...,n_f$ the charges of the $n_f$ massless quarks present in the plasma. The strong coupling  only enters in the quark current correlator $W^{\mu\nu}(P)$, which is calculated below at leading order but which receives large higher order corrections.
If we restrict to leading order and to the case where the lepton mass is negligible compared to the invariant mass $M=\sqrt{P^2}\gg m$, which is a good approximation for electrons, we get \cite{laine}:
\begin{equation}
\label{laineanalitic}
\frac{dR^{\ell \bar \ell}_{LO}}{d^{4}P}=\sum_{i=1}^{n_f} Q_i^2
\frac{\alpha_e^{2}}{2\pi^{4}}
\frac{T}{p} \frac{1}{e^{E/T}-1} \log{\frac{\cosh{\frac{E+p}{4T}}}{\cosh{\frac{E-p}{4T}}}},
\end{equation}
where $E=p_0$ and $p=p_i^2$.

\subsection{NLO corrections to the spectra}
\label{NLO}
\begin{figure}[H]
\begin{center}
\includegraphics[scale=0.095]{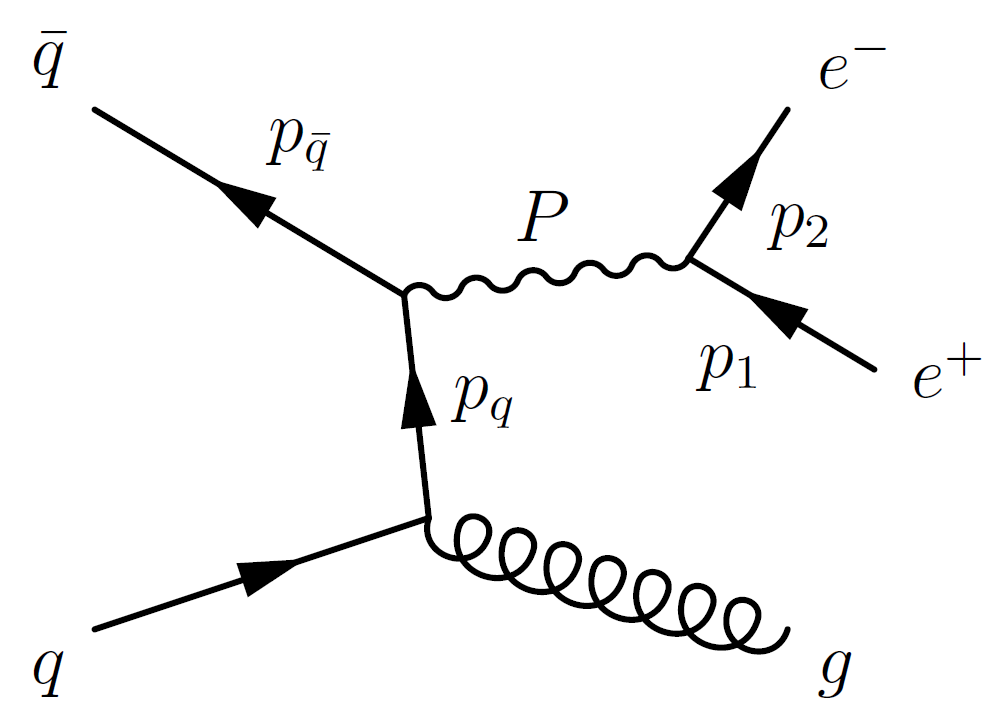}
\hspace{0.4cm}
\includegraphics[scale=0.095]{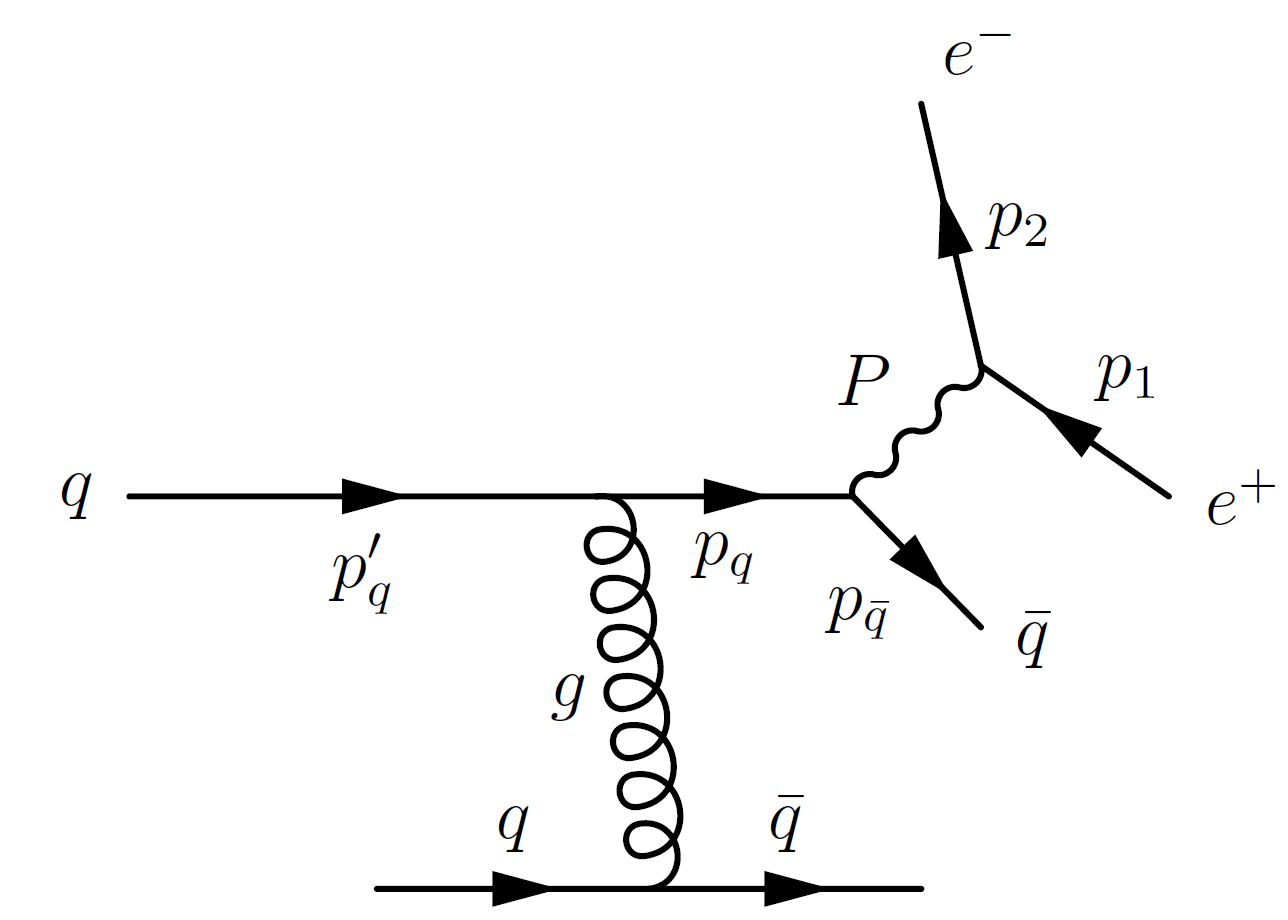}
\end{center}
\caption{Examples of NLO Feynman diagram for the thermal dilepton production from the QGP.}
\label{feyNLO}
\end{figure}

\begin{figure}[H]
\begin{center}
\includegraphics[scale=0.095]{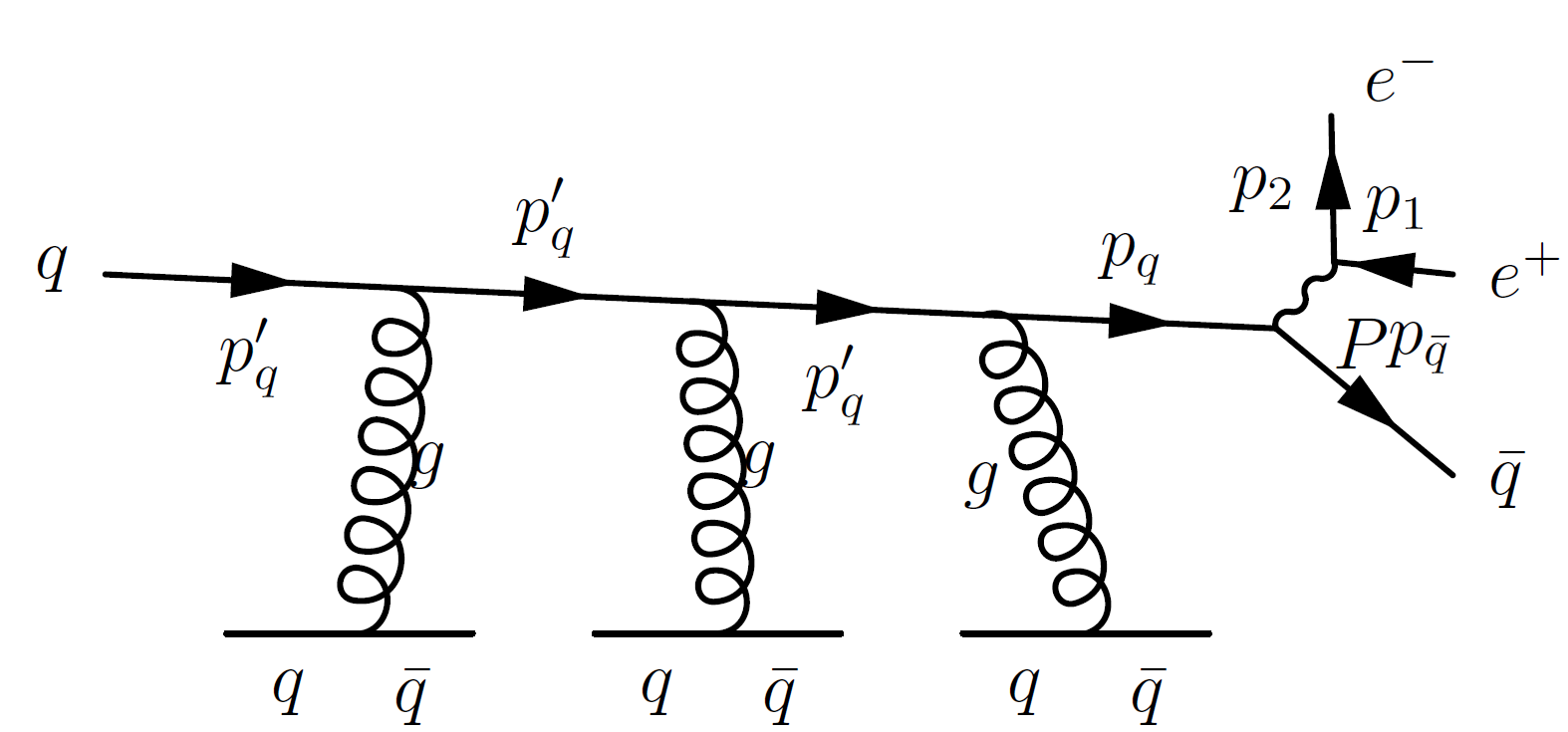}
\end{center}
\caption{Feynman diagram that summarizes the LPM effect: in the very dense QGP a quark is rescattered many times before it annihilates with its antiparticle.}
\label{feyLPM}
\end{figure}

As emphasised before, large corrections to the leading order dilepton rate arises. The next-to-leading order (NLO) is suppressed only by $\alpha_s$, but diverges in the small invariant mass limit $M\to 0$ (some representative diagrams for the NLO are shown in fig.~\ref{feyNLO}). In this soft limit the correct result can only be obtained by performing the Landau-Pomeranchuk-Migdal (LPM) resummation. The LPM resummation takes into account the destructive interference effect between the prompt emitted photons and it is summarised by the Feynman diagram in figure \ref{feyLPM}.


In this section we investigate how the NLO and LPM corrections contribute to the dilepton spectra. 
To show the results of NLO and higher order corrections, we used the data furnished by Ghisoiu and Laine \cite{laine} (NLO and LPM to LO) and Ghiglieri and Moore (for the LPM at NLO)  \cite{ghiglieri}. It consists in a database for the electron-positron and the muon-antimuon emission rate as a function of invariant mass $M$, temperature $T$ and modulus of 3-momentum $P$  for the NLO  and for the LPM corrections (available at \cite{laineweb}).

\begin{figure}[H]
\begin{center}
\includegraphics[scale=0.3]{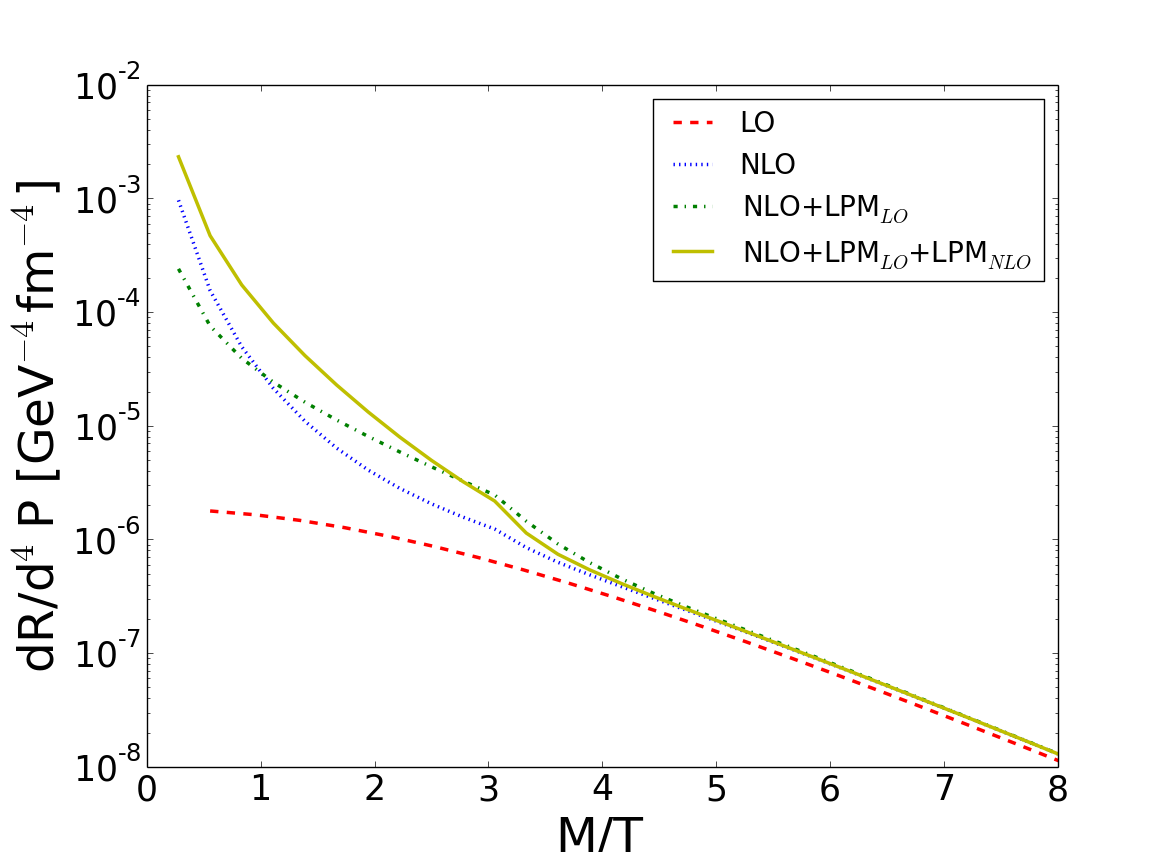}
\end{center}
\caption{dielectron emission rate computed at leading order, NLO \cite{laine} , and NLO plus LPM corrections  \cite{ghiglieri} as a function of the dimension-less quantity $M/T$ for the values $P=0.53$ GeV, $T=0.18$ GeV and the quarks $u,d,s$ have been considered. These corrections dominate at small total momentum $P$.
}
\label{NOvsNLO}
\end{figure}

Figure \ref{NOvsNLO} shows a comparison between the dielectron emission rate at LO computed with  formula (\ref{laineanalitic}) and the same with NLO and NLO plus LPM corrections. 

We can notice that the NLO and LPM contribution to the emission rate are important for small values of the ratio of the invariant mass over temperature $M/T$. Even if the LPM corrections suppress the rate at very small $M$, they increase the rate for intermediate values of the invariant mass.

\subsection{Geometry of HIC and hydrodynamics of the plasma}

Over all this work, we will keep separated the longitudinal dynamics (along the collision axis), from transverse dynamics, described by the (2+1)-dimensional hydrodynamic  model in section \ref{metho} \cite{rom1,rom2}.

We use the Bjorken model \cite{BJ,plateau} to describe the longitudinal expansion and thus we assume  the existence of a "central plateau" structure in the production rate of particle as a function of space-time rapidity
\begin{equation}
\zeta=\frac{1}{2}\ln\frac{t+z}{t-z}.\label{rapidity}
\end{equation}
Rapidity $\zeta$  and proper time $ \tau=\sqrt{t^2-z^2}$ are therefore a convenient reparametrization of $z$ and $t$ to describe the longitudinal flow:
\begin{equation}
\label{milne}
x^{\mu}=(\tau\cosh\zeta,\mathbf{x}_{\perp},\tau\sinh\zeta),
\end{equation}
so that one can rewrite the differential measure of space-time as $d^{4}X=\tau d\tau\, d\zeta\, d^{2}x_{\perp}$. 

The longitudinal Bjorken flow \cite{BJ} velocity is defined simply as the distance covered over proper time:
\begin{equation}
\label{bjvz}
u_{\mu}(\tau,y)=\gamma_{B}(1,0,0,v_z)=\gamma_{B}(1,0,0,\frac{z}{\tau}),
\end{equation} 
\noindent where $
\gamma_{B}=\frac{1}{\sqrt{1-(z/\tau)^2}}. $
On the other hand, we obtain the temperature in the $\zeta=0$ slice,  using a (2+1)-dimensional hydrodynamic simulation (see section \ref{metho}).

\subsection{Dilepton spectra in HIC}
\label{spectra}

One can rewrite formula (\ref{laineanalitic}) using a new parametrisation of the 4-momentum
of the virtual photon \cite{strick}, which is better suited for the geometry of HIC: 
\begin{equation}
\label{param}
p^{\mu}=\left(m_{\perp}\cosh y,p_{\perp}\cos\phi_{p},p_{\perp}\sin\phi_{p},m_{\perp}\sinh y\right).
\end{equation}
In the formula above, $\phi_{p}$ denote the  azimutal angle, the transverse mass is $m_{\perp}\equiv\sqrt{M^{2}+p_{\perp}^{2}}$ and $y$ the momentum space rapidity
\begin{equation}
y=\frac12\ln\frac{p^0+p^z}{p^0-p^z}.
\end{equation}
Using (\ref{param}), we  write
the differential 4-momentum as $d^{4}P=M\, dM\, dy\, p_{\perp}dp_{\perp}d\phi_{p}$.

We are now ready to compute the invariant mass and rapidity
differential spectra:
\begin{equation}
\label{15}
\frac{dN^{l^{+}l^{-}}}{M\, dM\, dy}=\int_{p_{\perp}^{min}}^{p_{\perp}^{max}}p_{\perp}dp_{\perp}\int_{0}^{2\pi}d\phi_{p}\int d^{4}X \frac{dR^{\ell \bar \ell}}{d^{4}P}\, .
\end{equation}
As we are interested only in the thermal contribution to the dilepton emission spectrum, the integration over
$d^{4}X$ is performed only on the quark-gluon plasma volume,
i.e. on the regions with
temperature bigger then the critical temperature $T({\bold x}_\perp,\tau)>T_{c}=0.17\, $ GeV.


It is important to note that, in formula (\ref{15}), the values of $p_{\perp}$ are defined in the LAB reference frame, but the emission rate $dR/d^4P$ (in formula (\ref{laineanalitic})) has been computed in local rest frame (LRF) of the plasma\cite{strick}. While integrating over the plasma volume, we have to boost the observed LAB frame momenta to the LRF of the plasma, which we can plug in the dilepton rate formula (\ref{laineanalitic}) computed previously.

\subsection{Change of frame}
\label{hystory}

Let us consider the integral over space-time  in (\ref{15}) and write it more explicitly using (\ref{milne}): $\int d^{4}X\,\frac{dR^{l^{+}l^{-}}}{d^{4}P}=\int \tau d\tau d{\bf x}_{\perp} d\zeta\,\frac{dR^{l^{+}l^{-}}}{d^{4}P}$. Notice that the emission rate (\ref{laineanalitic}) depends on ${\bf x}_{\perp}$ and $\tau$ only through the temperature $T({\bf x}_{\perp},\tau)$, which is given by the  hydrodynamical simulation, while the dependence on $\zeta$ is given by the Bjorken model, as anticipated. 

At the end of the previous subsection, we noticed that,  whenever we fix a LAB frame value of $p_{\mu}$ and a volume element in space in the integral (\ref{15}), it is necessary to boost the momentum to
the LRF of the volume element of which we want to compute the emission rate.

%
%
In order to do this, we need the boost $\Lambda^{\mu}_{\nu}(u^{\mu})$  that parametrizes the change of coordinates from the LAB frame to the LRF \cite{strick}. It is defined by the relative velocity  $u^{\mu}_{tot}(x^{\mu})$ that combines the Bjorken longitudinal  velocity with the transverse hydrodynamical expansion.
Once we found $u^{\mu}_{tot}(x^{\mu})$,  we are able to compute (\ref{15}) considering $\frac{dR^{l^{+}l^{-}}}{d^{4}P}\left(\Lambda^{\mu}_{\nu}(u_{tot})p^{\nu}_{LAB}\right)$. In what follows, we show the derivation of $u^{\mu}_{tot}(x^{\mu})$.


From the hydrodynamical simulations  we obtain the 4-velocity of the transverse flow (for $\zeta=0$), 
$$u^{\mu}_{hydro}=\gamma_{hydro} \{ 1, v^{hydro}_x,v^{hydro}_y,0\} ,$$
where  $\gamma_{hydro}=\sqrt{1+(u^{\mu }_{hydro})^{2}}$ and $v^{hydro}_x$ and $v^{hydro}_y$ are measured with respect to the collision axis. However the whole system is moving along the longitudinal axis with Bjorken velocity $v_z$, estimated  in (\ref{bjvz}).
Thus we will need to use the relativistic composition law to find the total velocity respect to the LAB frame.

\begin{figure}[H]
\begin{center}
\vspace{-3mm}
\includegraphics[scale=0.22]{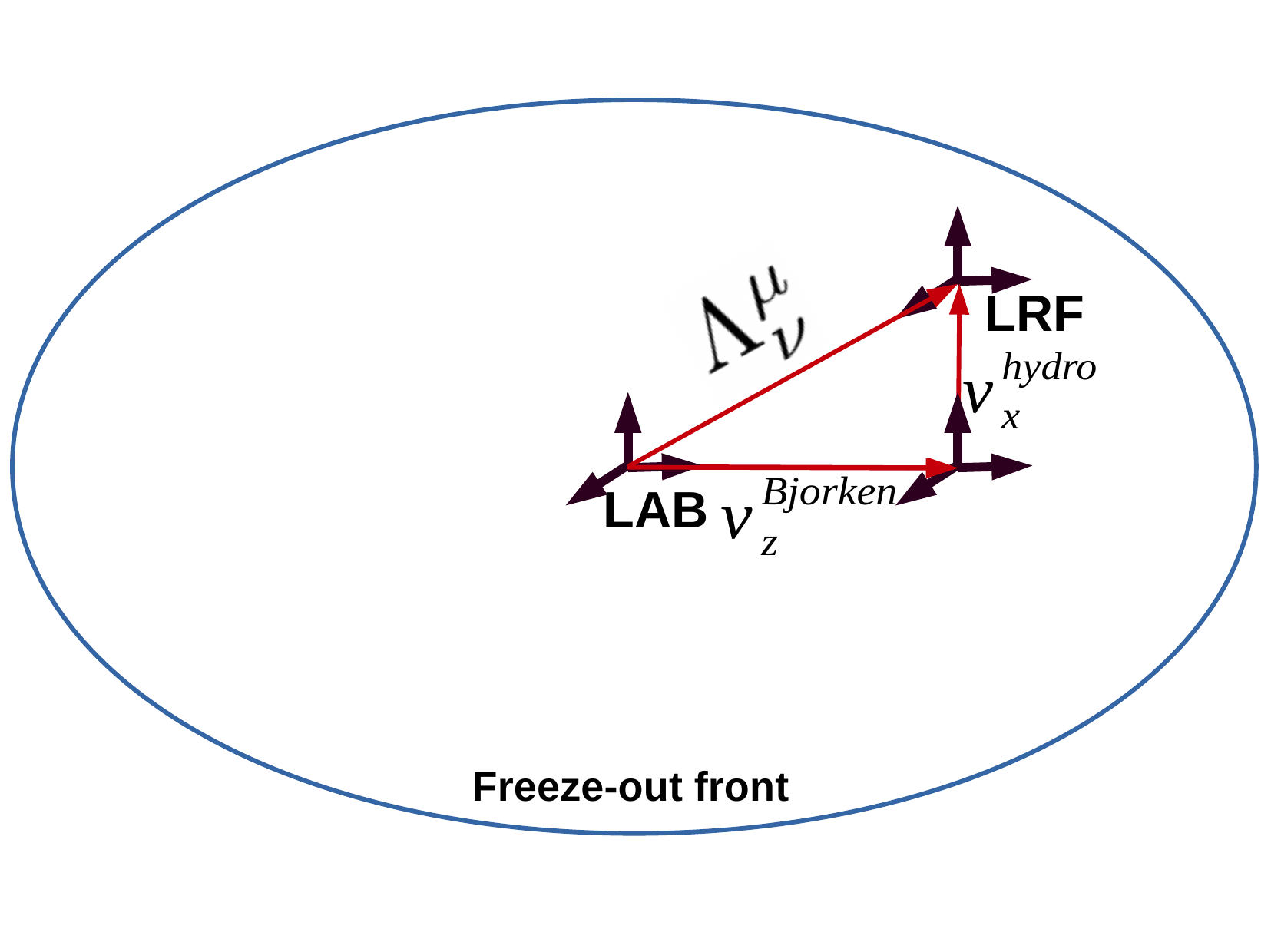}
\vspace{-3mm}
\end{center}
\caption{Scheme of the combination of longitudinal and transverse velocity, and the final boost $\Lambda^{\mu}_{\nu}$. }
\label{lambda}
\end{figure}

\noindent The total velocity of a generic element of volume inside the QGP is the relativistic sum of $v_z$ and $\textbf{v}^{hydro}$:  $v_x= v_z\oplus v^{hydro}_x  $ and $v_y=v_z\oplus v^{hydro}_y$.
\footnote{The sum of relativistic velocities is not commutative. According to Bjorken model, we make the hypothesis that, in the central rapidity plateau, the transverse evolution of the plasma is the same at any rapidity. Thus a generic slice of plasma at a generic rapidity $\zeta_0$ evolves radially exactly like the $\zeta_0=0$ slice described by hydro model. We want to obtain the total velocity of a generic element of volume of the plasma respect to the LAB frame where $p^{\mu}$ is measured. The LAB frame, in our case, corresponds to the CM frame, thus every time we fix $p^{\mu}$ in the LAB frame we first have to make a boost along the longitudinal axis to the center of the "$\zeta_0$-plasma slice"  and then add the transverse velocity relative to that point, obtained with the hydrodynamics simulations. }

 Then, applying the relativistic  addition rule for perpendicular velocities $\textbf{ v}_{tot}=\textbf{v}_1+\sqrt{1-v_1^2}\textbf{v}_2$, one obtains
$
u_{tot}(x)=\gamma_{tot}\left(1,\gamma_B^{-1} v_x^{hydro},\gamma_B^{-1} x_y^{hydro}, \frac{z}{t}\right)
$, where  $
\gamma_{tot}=
\gamma_{Bj}\cdot \gamma_{hydro}$.

\section{Numerical simulations}
\label{num}
\subsection{SONIC}
\label{metho}

The collisions between two heavy ions are not well understood at early times, before the system thermalizes. As soon as the QGP is formed (at proper time $\tau_{in} \sim 0.5$ fm),  its space time evolution  is described by hydrodynamic models \cite{artstory}. 

We simulate the hydrodynamic evolution of the QGP using the software SONIC (Super hybrid mOdel simulatioN for relativistic heavy-Ion Collisions), developed by Romatschke, Luzum and others (the code is available at \cite{site}) \cite{rom1,rom2,rom3,rom4,rom5,rom6,rom7,rom8}.
In this section we summarize the model on which SONIC is based on.

It consists into a (2+1)-dimensional model that takes into account  only  the slice at rapidity $\zeta=0$, in which   the center of mass lies. 
It combines the pre-equilibrium flow, modelled as in Ref.~\cite{rom1}, with the hydrodynamic phase and the latter with the final hadronisation (which does not concern this work).

SONIC simulates the  highly boosted and Lorentz contracted nuclei starting with their energy density, $T_{tt}=\delta(t+z)T_{A}({\bold x}_\perp)$, where the function
$T_{A}$ has the shape \cite{rom2}:
\begin{equation}
\label{overlap}
T_{A}=\epsilon_{0}\int_{-\infty}^{\infty}dz\left[1+e^{-(\sqrt{{\bold x}_\perp^{2}+z^{2}}-R)/a}\right]^{-1},
\end{equation}

\noindent $R$ and $a$ are the charge radius and the skin depth parameters
(the values of these parameters can be found in table 1 in reference \cite{rom2}) and $\epsilon_{0}$ is a normalization constant that
controls the final charged multiplicity and it is set to reproduce the available experimental data 

The pre-equilibrium radial flow velocity is estimated numerically in \cite{rom1}:
\begin{equation}
\label{vel}
v^{\perp}_i(\tau,{\bold x}_\perp)=-\frac{\tau}{3.0}\partial_{i}\ln T_{A}^{2}({\bold x}_\perp),
\end{equation}

\noindent where $\tau=\sqrt{t^{2}-z^{2}}$.
The initial energy profile is set to be 
\begin{equation}
\label{energy}
\epsilon(\tau,{\bold x}_\perp)=T_{A}^{2}({\bold x}_\perp).
\end{equation}

SONIC  includes the relativistic viscous hydrodynamics
solver, (here we use version 1.7), that implements the evolution of the system using the energy density from equation (\ref{energy}) and the flow profile from equation (\ref{vel}).


The main  parameters that need to be set in the hydrodynamical simulation are: the freeze-out temperature $T_{c}=$0.17 GeV; the initial central temperature $T=$0.37 GeV for RHIC and $T=$0.47 GeV for LHC; the shear viscosity $\eta/s=1/4\pi$.

\subsection{Integration of the dilepton rate}

In this subsection, we  enunciate the details of the numerical computation for the dilepton spectra (\ref{15}).

Firstly we must introduce the setting to the SONIC simulations. 
The starting proper time is $\tau_{start}=0.5$ fm and the temporal lattice spacing is 0.001 fm. 
The space grid (which spans the x-y plane) is made of 139  lattice sites (per each dimension), separated by $dx=dy=1$ GeV$^{-1}$, it covers the squared area $[-13.6,13.6]^2$ fm$^2$.
Every 500 time steps, (0.5 fm), it takes a "snapshot", i.e. it writes into data files all the measurables, from which we use the temperature and the transverse velocity.

From the inner to outer integration in (\ref{15}), we first computed the integration over $\tau d\tau d\zeta$ with the method of parallelepipeds. We integrated $\zeta$ in the half range $[0,0.9]$ divided in 20 steps and then double the result (the integral is symmetric for positive and negative values of $\zeta$).
The other integrals (over the transverse coordinates $x$ and $y$) are computed separately with the method of trapezes on the same lattice than the SONIC simulation. 

The integral $\int d\phi_P$ is computed in $[0,\pi/2]$ with 4 steps and then we multiply the result times 4, (the system is symmetric under rotation with period $\pi/2$).
The extremes of integration for $p_{\perp}$ for RHIC simulations have been chosen as in STAR: $p_{\perp}\in[0.2,15]$ GeV  (we note that contributions from $p_{\perp}>15$ are negligible), while we choose $p_{\perp}\in[0,15]$ GeV for LHC and the integral is computed in 34 steps. 

The vales of $\frac{dR^{l^{+}l^{-}}}{d^{4}P}$ were tabulated in advance as a function of $T$, $M$ and $p_{\perp}$ and the values of $M$ that we plot always correspond to nodes of the 3D mesh on which  $\frac{dR^{l^{+}l^{-}}}{d^{4}P}$ is tabulated. 

The same was not possible for $p_{\perp}$ because the boost  shifts its value. Thus for a give $p_{\perp}$ and $T$ we find the corresponding  $\frac{dR^{l^{+}l^{-}}}{d^{4}P}$ by bi-linear interpolation \cite{rom2}.



\section{Results}
\label{res}


\subsection{Predictions for RHIC}

\begin{figure}[H]
\begin{center}
\includegraphics[scale=0.27]{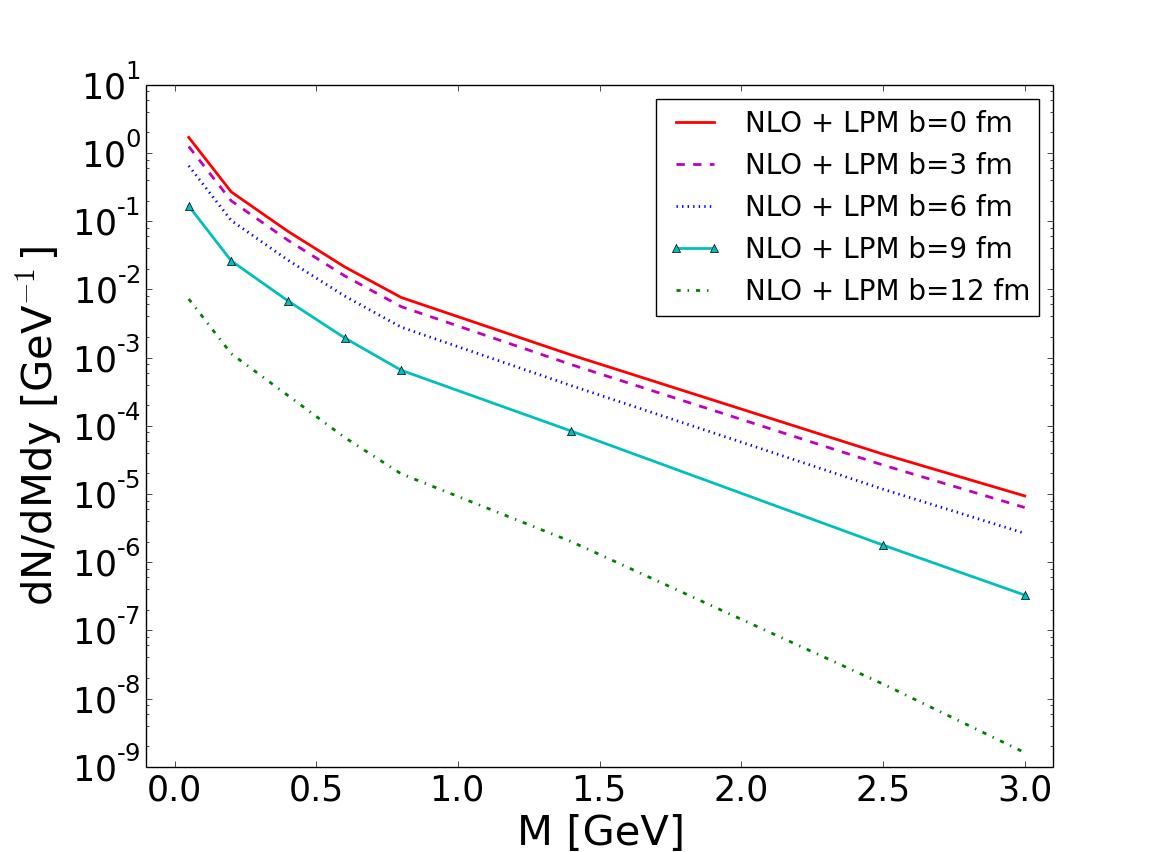}
\end{center}
\caption{Invariant mass spectra for the dileptons emission at RHIC computed at NLO with LPM corrections for different values of the impact parameter b, for central rapidity $y=0$, and transverse momentum  $p_{\perp}\in[0.2,15]$ GeV.}
\label{rhicbs}
\end{figure}

In this section, we present the results for the mass dielectron spectrum of the quark gluon plasma  (\ref{15}) for Au-Au collisions at $\sqrt{s}=200$ GeV  RHIC and for Pb-Pb collisions at $\sqrt{s}=2.76$ TeV at LHC. Moreover we compare the RHIC results with the experimental data from the STAR experiment. 

Figure \ref{rhicbs} shows our results for the thermal dilepton spectrum (\ref{15}) as function of invariant mass $M$, for different values of impact parameter $b$ and for LO, NLO and NLO + LPM approximation. 

STAR measures the electron-positron pairs from Au-Au ions collisions at  $\sqrt{s}=200$ GeV, as a function of the invariant mass $M$ of the virtual photon  and its transverse momentum $p_{\perp}$ \cite{star}. STAR can capture emitted leptons at all azimuthal angles and for momentum-space rapidity values $\abs{y}<1$. 
To make a good comparison to experiment, we also integrate the dilepton spectrum over momentum-space rapidity $\abs{y}<1$.

\begin{figure}[H]
\begin{center}
\includegraphics[scale=0.27]{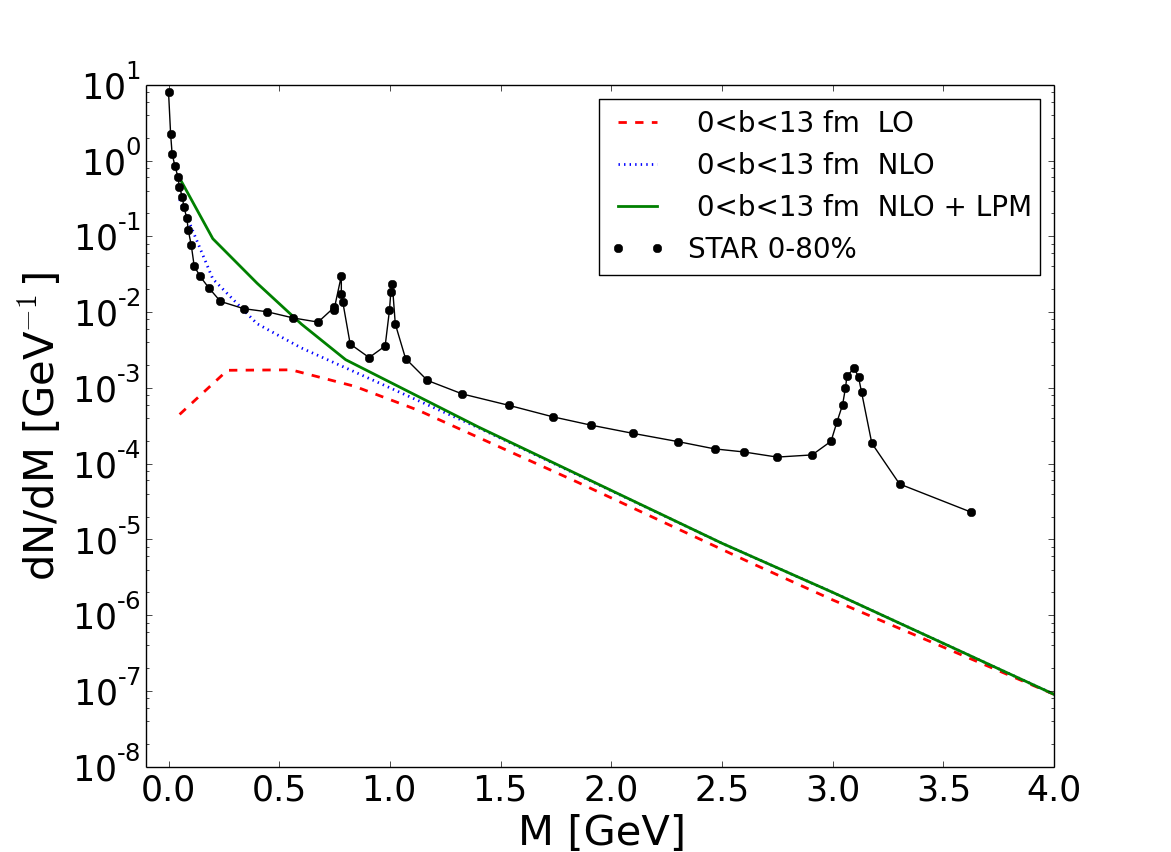}

\caption{Invariant mass spectrum for the thermal emitted dileptons computed at LO, NLO and NLO+LPM corrections for the full detected centrality range (0-80\%) and comparison with the relative data from STAR. Error bars for the star data can be found in \cite{star}.}
\label{rhic080}
\end{center}
\end{figure}

Moreover, the data are classified by  centrality ranges, thus it was necessary to average our results over the impact parameter $b$ in order to reproduce the centrality ranges.  To this aim, we integrated over different impact parameters b as  \footnote{The integral over $y$ and the one at the numerator of (\ref{inte_b}) has been computed with Simpson's rule for parabolic integration. Considering the dependence of the emission rate on the impact parameter (see references \cite{cetrality_rhic,brhic}), this should give the exact solution of the integral in the range $b\in[0,13]$ fm. }
\begin{equation}
\frac{dN}{dM}(\%centrality)=\frac{\int_{b_{min}}^{b_{max}}db\,b\frac{dN}{dM}(b)}{\int_{b_{min}}^{b_{max}}db\,b}
\label{inte_b}
\end{equation}
\noindent where $b_{min}$ and $b_{max}$ can be found as a function of centrality in \cite{cetrality_rhic,brhic} \footnote{Analogue tables are  in ref. \cite{blhc} for LHC. }.
Figure \ref{rhic080} shows the dilepton spectrum for RHIC averaged on all the centralities 0-80\%, which, for Au-Au collisions, corresponds to $b=0-13$ fm, for different orders in perturbation theory.

Of course, the experimental data from \cite{star} include, in addition to the thermal dileptons, high energetic electron-positron pairs from pre-equilibrium processes and mainly the ones generated in  the following decays: $\omega \rightarrow e^+ e^- \pi ^{0}$,
 $\pi^0 \rightarrow e^+ e^- \gamma$, 
  $\eta \rightarrow e^+ e^- \gamma$, $\eta^0$,
   $\omega \rightarrow e^+ e^-$, $,\rho  \rightarrow e^+ e^-$
  $\phi \rightarrow e^+ e^-$,
  $J/\psi \rightarrow e^+ e^- X$ .
For large invariant mass, the thermal contribution quickly becomes small in comparison to the other contributions discussed in the introduction \cite{contributions,sources,Vujanovic:2012nq,Vujanovic:2013jpa}. 
In the high mass range $2<M<3$ GeV, the contribution from particle decays is much more important then the thermal one. For $M>3$ GeV the main contribution to the spectrum is given by dielectrons couples produced in pre-equilibrium Drell-Yan processes, and our forecast are, of course, much smaller then experimental data.

The region in which the thermal contribution is dominant is indeed very small, i.e. roughly $0.2<M<1.5$ GeV, up to the $\rho, \omega$ and $ \phi$ peaks. In this region the agreement can be tested.

\begin{figure}[H]
\includegraphics[scale=0.27]{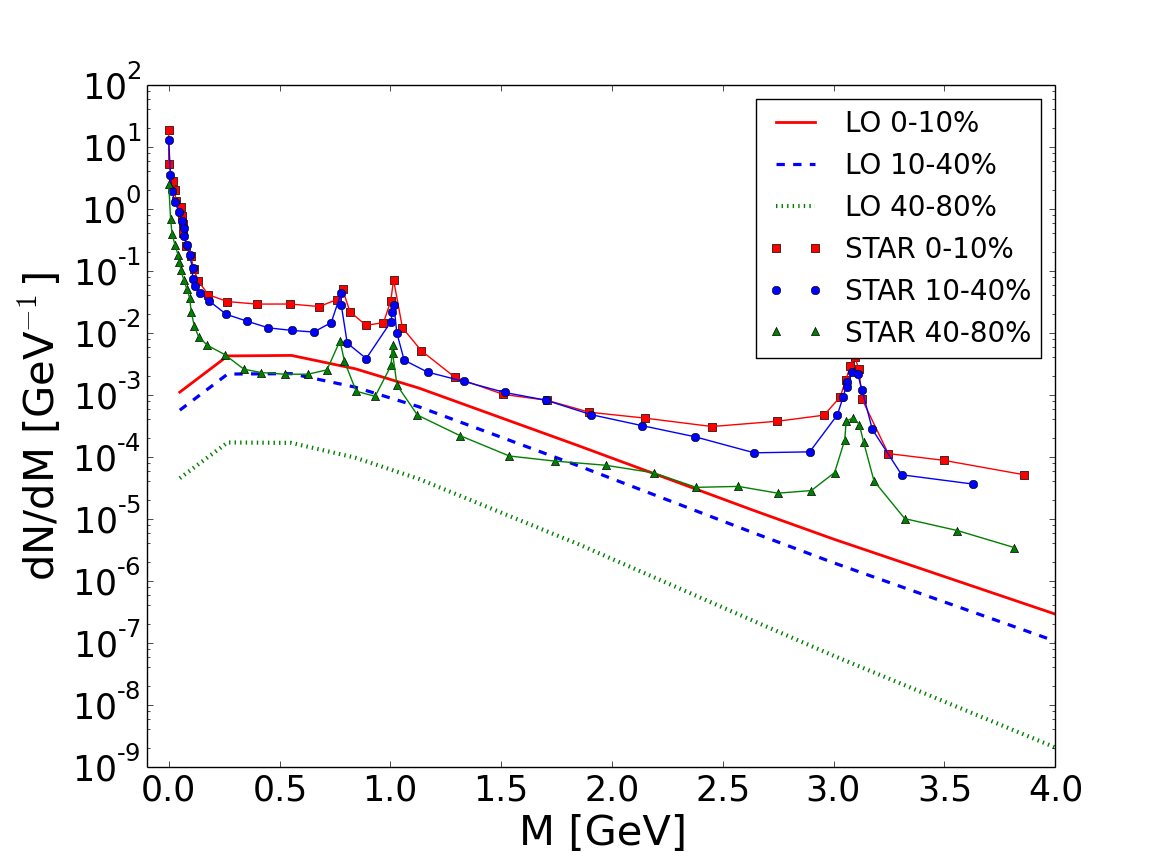}\\
\includegraphics[scale=0.27]{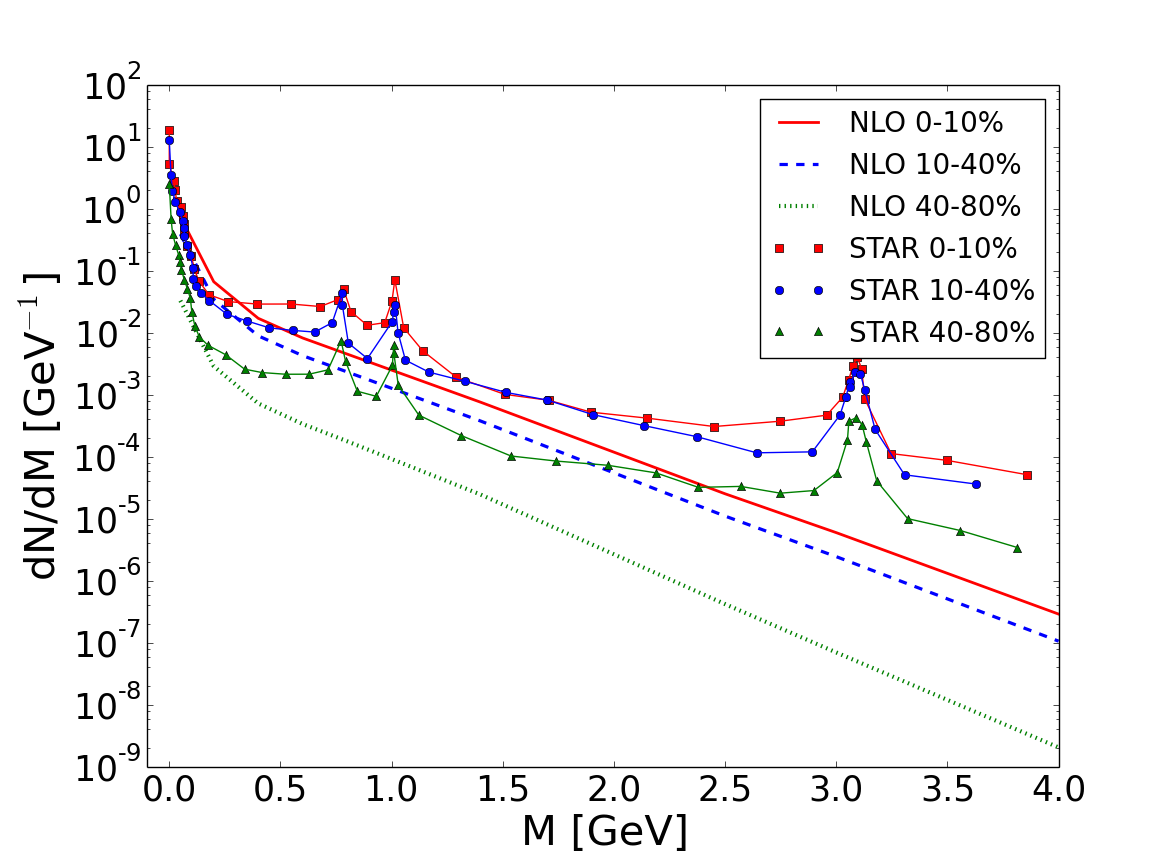}\\
\includegraphics[scale=0.27]{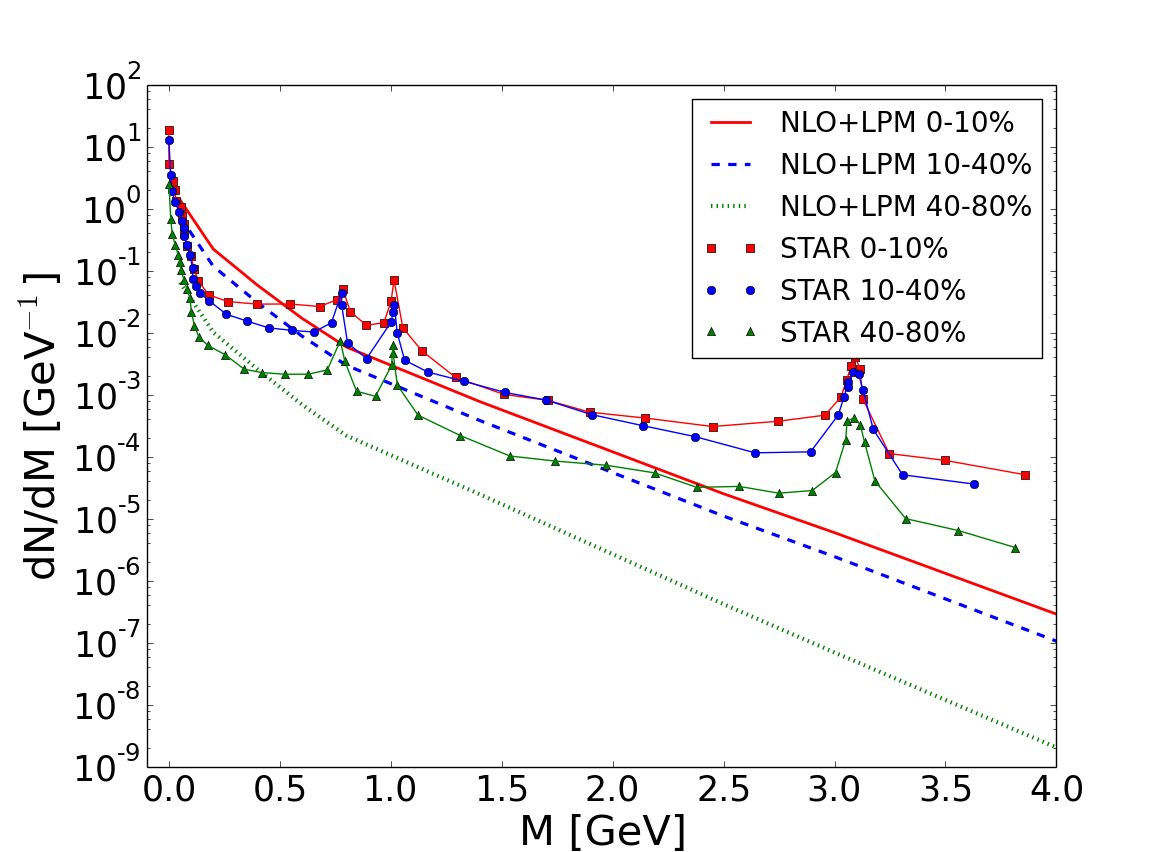}\\
\caption{From up to down: Invariant mass spectra for the dileptons emission at RHIC computed at LO, NLO and NLO with LPM corrections respectively, for different ranges of centrality. In each plot the predictions based on (\ref{15}) are compared with the relative datas from STAR. The predicted spectra consider the rapidity range $\abs{y}<1$ and transverse momentum range $p_{\perp}\in[0.2,15]$ GeV.}
\label{rhic3}
\end{figure}

Figure \ref{rhic3} shows the comparison between our predictions and the experimental data for different range of centrality, computed as in (\ref{inte_b}).  Surprisingly, the LPM corrections overestimate the number of emitted thermal dileptons at small invariant mass and the NLO approximation is the closest to experimental data. 
In the very low mass range $M<0.5$ GeV, perturbation theory breaks down, and a different approach should be studied, for example lattice simulations.

For $M>0.7$ GeV, our results are a bit smaller than experimental data as expected, since we do not consider all the contributions that are included to the experimental data. Moreover we can notice that, the more the impact parameter $b$ increases, the farthest are the predictions with the experimental data. This is expected, as for large $b$, the volume of the plasma produced is smaller and so the thermal contribution to the invariant mass dilepton spectrum is less important.

\subsection{Predictions for LHC}

\begin{figure}[H]
\includegraphics[scale=0.27]{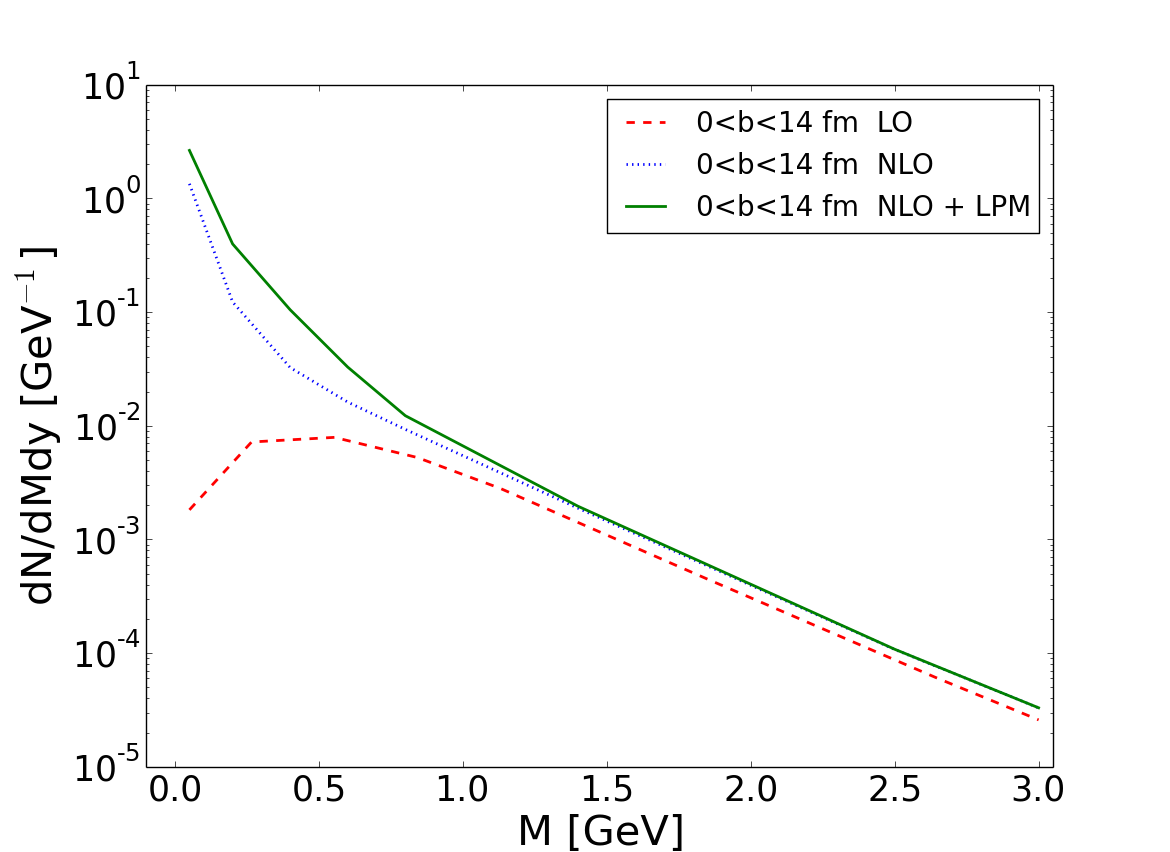}
\caption{Invariant mass spectra for the dileptons emission at LHC computed at LO, NLO and NLO with LPM corrections for collisions with centrality in range 0-80\%. The curves are computed at mid-rapidity $y=0$, for the transverse momentum range: $p_{\perp}\in[0,15]$ GeV.}
\label{lhc080}
\end{figure}

Figure \ref{lhc080} is the analogue of figure \ref{rhic080} but for Pb-Pb collisions at $\sqrt{s}=2.71$ TeV at LHC, with the difference that we kept the rapidity $y=0$ and we integrated over the full transverse momentum range. The number of emitted dielectrons pairs is of course bigger. 

\begin{figure}[H]
\includegraphics[scale=0.27]{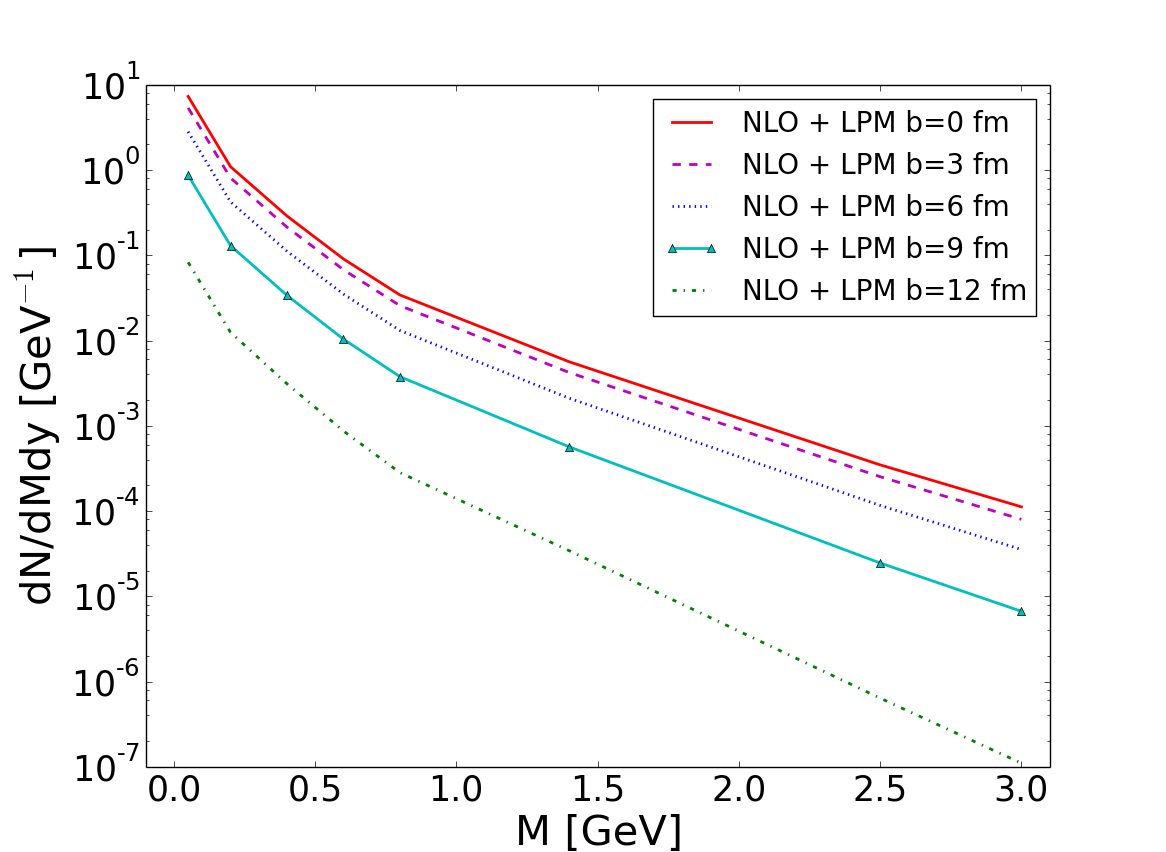}\\
\caption{Invariant mass spectra for the dileptons emission at LHC computed at  NLO with LPM corrections for different values of the impact parameter b. The curves are computed at mid-rapidity $y=0$, for the transverse momentum range: $p_{\perp}\in[0,15]$ GeV.}
\label{lhcbs}
\end{figure}

\begin{figure}[H]
\includegraphics[scale=0.27]{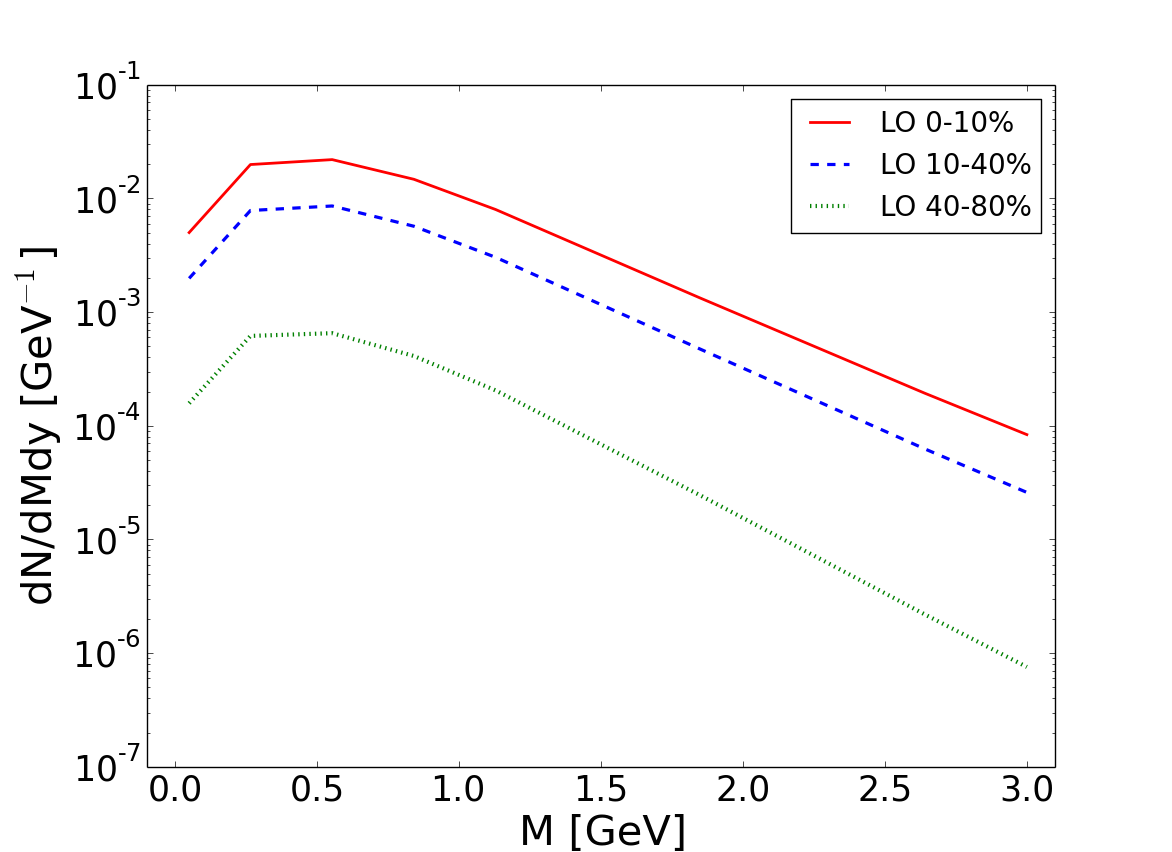}\\
\includegraphics[scale=0.27]{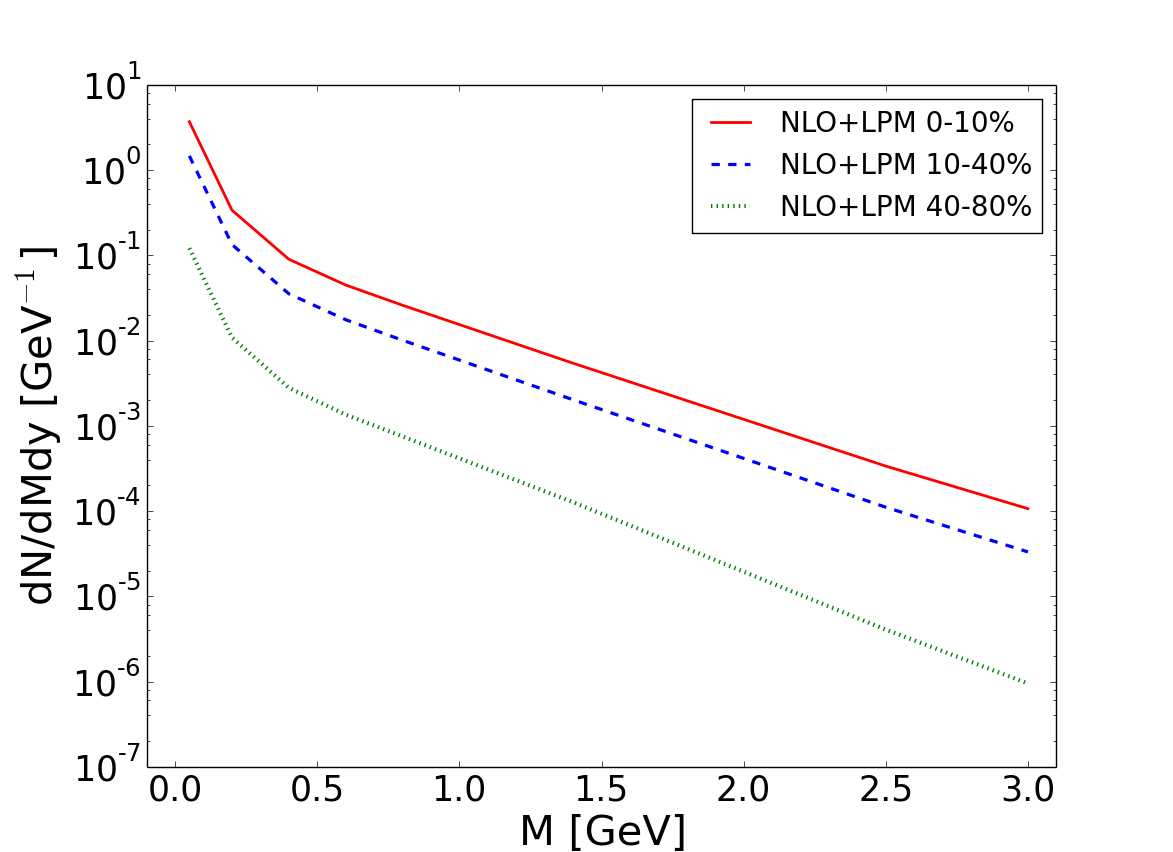}\\
\includegraphics[scale=0.27]{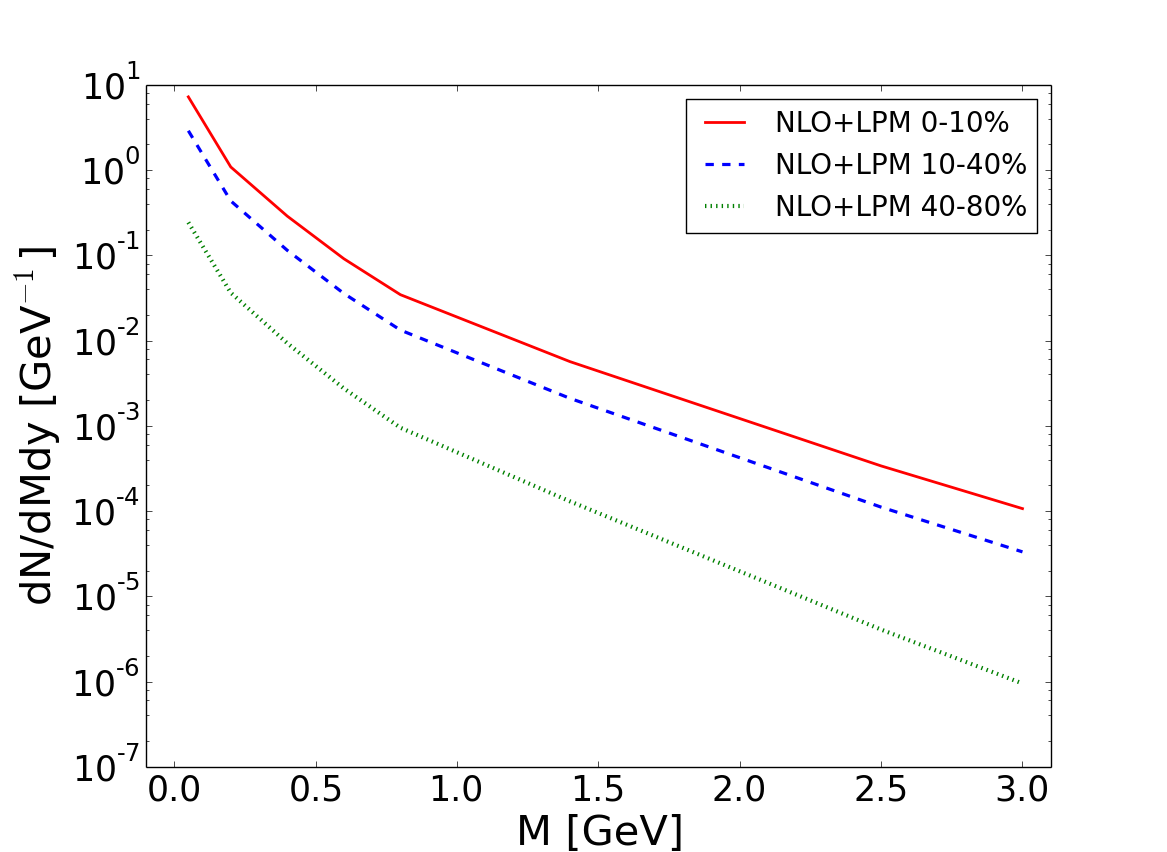}
\caption{From up to down: invariant mass spectra for the dileptons emission at LHC computed at LO, NLO  and NLO +LPM corrections for different ranges of centrality. The curves are computed at mid-rapidity $y=0$, for the transverse momentum range: $p_{\perp}\in[0,15]$ GeV.}
\label{lhc3}
\end{figure}

Figure \ref{lhcbs} and figure \ref{lhc3} are the analogues of figure \ref{rhicbs}.3 (NLO + LPM corrections) and \ref{rhic3}.2 (NLO) respectively.

\section{Conclusion}
\label{concl}

The main result of this work is the comparison between the thermal dilepton rate calculated at LO, NLO and NLO+LPM.  The higher order corrections becomes important for small value of the invariant mass of the virtual photon $M$. Comparing our results with experimental data from the STAR experiment at RHIC, we see that for small invariant mass, the NLO+LPM rate seems to overshoot the data. This shows that the LPM effect, even if it damps the rate at very small $M$ actually enhance the rate for $M\sim0.5$GeV too much to be compatible with the STAR experiment. The NLO seems to fit experiment best but overshoot the data for $M<0.5$ GeV. For such small values of $M$ it seems that a non-perturbative determination is a must. 

We also performed predictions for the LHC, where the plasma phase might become more important in comparison to other sources. Results from LHC would be very useful to settle the tension between STAR data and higher order calculations of the dilepton rate.

\end{multicols}
\end{document}